\documentstyle[12pt]{article}

\newcommand{\beq}{\begin{equation}}
\newcommand{\eeq}{\end{equation}}
\newcommand{\beqn}{\begin{eqnarray}}
\newcommand{\eeqn}{\end{eqnarray}}
\newcommand{\dif}{\partial}

\title{
 The Boundary State Formalism
and Conformal Invariance in Off-shell String Theory }

\author{ M. Laidlaw and  G. W. Semenoff
\\~~\\
Department of Physics and Astronomy, \\
 University of British Columbia, \\
Vancouver, British Columbia, Canada V6T 1Z1.
}

\begin{document}
\bibliographystyle{unsrt}

\maketitle

\abstract{
In this note we present a generalization of the boundary state formalism for the
bosonic string that allows us to 
calculate
the overlap of the boundary state with arbitrary closed string states.
We show that this generalization exactly reproduces world-sheet sigma model calculations, 
thus giving the correct overlap with both on- and off-shell string states, and 
that this new boundary 
state automatically satisfies the requirement for
integrated vertex operators in the case of non-conformally invariant boundary interactions}

\newpage

\section{Introduction}

The problem of studying off-shell string theory is an old one, and there have been many
attempts to examine it.  A particularly interesting approach is
 background independent string 
field theory 
\cite{Witten:1992qy,Witten:1993cr,Witten:1993ed,Shatashvili:1993kk,Shatashvili:1993ps}
which has received attention recently as an approach which facilitates the understanding
of properties of unstable d-branes \cite{Gerasimov:2000zp,
Gerasimov:2000ga,Kutasov:2000qp}.
A tractable problem that can be approached within this formalism is to understand
the behavior of the off-shell theory in the background of a tachyon field, in particular
a quadratic function of the coordinates.
This model has been the subject of some research interest
\cite{Gerasimov:2000zp,
Gerasimov:2000ga,Kutasov:2000qp,Akhmedov:2001jq,Gerasimov:2001pg,Kraus:2000nj,
Craps:2001jp, 
Viswanathan:2001cs,Rashkov:2001pu,Alishahiha:2001tg,Andreev:2000yn,Arutyunov:2001nz,
deAlwis:2001hi}
and the resulting theory is not invariant under conformal transformations
except in the trivial cases of vanishing or infinite quadratic tachyon
potential. 
This subtlety reveals an interesting structure.
In this note we suggest a generalization of the boundary state \cite{Bardakci:2001ck,
Lee:2001cs,Lee:2001ey,Fujii:2001qp,DiVecchia:1999fx,Akhmedov:2001yh}
which naturally accommodates this loss of invariance.  This new boundary
state can be applied
to the problem of computing off-shell closed string emission amplitudes from a
d-brane.

For definiteness, we consider the following action on the string world sheet
\beqn
S(g,F,T_0,U) &=& \frac{1}{4 \pi \alpha'} 
\int_\Sigma d^2\sigma g_{\mu\nu} \dif^a X^\mu \dif_a X^\nu  
\nonumber \\
&~& + \int_{\dif \Sigma} d\theta \left( \frac{1}{2} F_{\mu\nu} X^\nu \dif_t X^\mu
+ \frac{1}{2\pi} T_0 + \frac{1}{8 \pi} U_{\mu\nu} X^\mu X^\nu \right). 
\label{action}
\eeqn
Here, $\alpha'$ is the inverse string tension, $\Sigma$ is the string world sheet, $d^2\sigma$ is
the measure on the bulk, $d\theta$ is the measure on the boundary, and $\dif_t$ is the
tangential derivative to the boundary.
The background fields that are included in this are $F_{\mu\nu}$, a constant gauge field strength,
and $T(X) = \frac{1}{2\pi} T_0 + \frac{1}{8 \pi} U_{\mu\nu} X^\mu X^\nu$, the tachyon profile
parameterized by a constant and a symmetric matrix.

We wish to show that the boundary state that we propose will provide an algebraic method
to calculate results that could be obtained in the $\sigma$-model calculations, and so
we will be comparing the results of $\sigma$-model calculation in these backgrounds
with analogous results 
obtained through boundary state calculations.
This will fix the normalization of
the boundary state 
and verify that it gives the results expected in the background of
the tachyon condensate.

\section{Sigma Model calculations}

It is instructive to commence by calculating some closed string emission amplitudes from
the disk world-sheet, since the boundary state will be seen to interact with closed string
modes in this way.  
First, it is useful to fix some conventions, the functional integral which we 
compute is in all cases an average over the action given in (\ref{action}),
\beq
\langle {\cal O}(X) \rangle = \int {\cal D} X e^{-S(X)}  {\cal O}(X). 
\eeq
In addition, the greens function on the unit disk with Neumann boundary conditions 
is determined to be \cite{Hsue:1970ra}
\beq
G^{\mu\nu}(z,z') = - \alpha' g^{\mu\nu} \left( -  \ln \left| z-z' \right| + \ln
\left| 1- z \bar z' \right| \right),  
\label{diskprop}
\eeq
and it will be useful also to know the bulk to boundary propagator
which is
\beq
G^{\mu\nu}(\rho e^{i \phi}, e^{i \phi'} ) = 2 \alpha' g^{\mu\nu}
\sum_{m=1}^\infty \frac{\rho^m}{m} 
\cos[ m(\phi - \phi') ].
\label{diskbbprop}
\eeq
The boundary to boundary propagator can be read off from (\ref{diskbbprop}) as
the limit in which $\rho \rightarrow 1$.
Throughout, we will use $z=\rho e^{i \phi}$ as a parameterization of the points within
the unit disk, so $0 \leq \rho \leq 1$ and $0 \leq \phi <  2\pi$.
Using the bulk to boundary propagator it is possible to integrate out the quadratic 
interactions on the boundary \cite{Fradkin:1985qd}
and to obtain an exact propagator, which is given by
\beqn
G^{\mu\nu} (z,z') &=& \alpha' g^{\mu\nu} \ln \left| z - z' \right| - \alpha' g^{\mu\nu} 
\ln \left| 1 - z \bar z' \right| \nonumber \\
&~& - \alpha' \sum_{n=1}^\infty \left( \frac{ 2 \pi \alpha' F + \frac{ \alpha'}{2}
\frac{U}{n} }{ g + 2 \pi \alpha' F + \frac{ \alpha'}{2} \frac{U}{n} } \right)^{\mu\nu}
\frac{(z \bar z')^n + (\bar z z')^n}{n} \nonumber \\
&=& \alpha' g^{\mu\nu} \ln \left| z - z' \right|
+ \frac{\alpha'}{2}
\sum_{n=1}^\infty \left( \frac{g - 2 \pi \alpha' F - \frac{ \alpha'}{2}
\frac{U}{n} }{ g + 2 \pi \alpha' F + \frac{ \alpha'}{2} \frac{U}{n} } \right)^{\mu\nu}
\frac{(z \bar z')^n + (\bar z z')^n}{n}.
\nonumber \\
\eeqn
 
The first interesting calculation that can be done is the partition function, which
has been calculated in several ways in the literature.  In the $\sigma$-model approach
the oscillator modes of $X$ must be integrated out with the contributions from $F$ and $U$
treated as perturbations.  Since both perturbations are quadratic, all the feynmann graphs
that contribute to the free energy can be written and evaluated, and explicitly (using the
parameterization $z= \rho e^{i \phi}$) the free energy is given by the sum
\beqn
{\cal F} &=& \sum_{n=1}^\infty \frac{1}{n} \int d\phi_1 \ldots d\phi_n  (-1)^n\left[ \left(
F_{\mu_1 \nu_1} 
\dif_{\phi_1} + \frac{1}{4\pi} U_{\mu_1 \nu_1} \right) \times
\right. \nonumber \\
&~& \left. 2 g^{\nu_1 \mu_2}
\sum_{m_1 = 1}^\infty \frac{ \cos[ m_1 (\phi_1 - \phi_2 ) ] }{m_1} \ldots 
 \left(
F_{\mu_n \nu_n}
\dif_{\phi_n} + \frac{1}{4\pi} U_{\mu_n \nu_n} \right) 2 g^{\nu_n \mu_1} \times
\right. \nonumber \\
&~& \left. 
\sum_{m_n = 1}^\infty \frac{ \cos[ m_n (\phi_n - \phi_1 ) ] }{m_n} \right] 
\nonumber \\
&=& - \sum_{m=1}^\infty Tr \ln \left( g + 2\pi \alpha' F + \frac{\alpha'}{2} \frac{U}{m} 
\right),
\label{messyeqn}
\eeqn
see \cite{Fradkin:1985qd,Laidlaw:2000kb} for further calculations done in this spirit.
From 
(\ref{messyeqn}) 
we immediately see that the partition function is given by
\beqn	
Z &=& e^{-T_0} \prod_{m=1}^\infty \frac{1}{\det \left(
g + 2\pi \alpha' F + \frac{\alpha'}{2} \frac{U}{m} \right) }
\int dx_0 e^{- \frac{ U_{\mu\nu} }{4} x_0^\mu x_0^\nu }  
\nonumber \\
&=& \frac{1 }{\det \left( \frac{U}{2}  \right) }  e^{-T_0} \prod_{m=1}^\infty \frac{1}{\det \left(
g + 2\pi \alpha' F + \frac{\alpha'}{2} \frac{U}{m} \right) }.
\eeqn
This expression is divergent, but using $\zeta$-function regularization 
\cite{Kraus:2000nj} it
can be reduced to
\beq
Z = e^{-T_0} \sqrt{ \det\left( \frac{g + 2\pi \alpha' F}{U/2 } \right) }
\det \Gamma \left( 1 + \frac{\alpha' U /2}{g + 2\pi \alpha' F} \right),
\label{partitionfunction}
\eeq
where $\Gamma(g)$ is the $\Gamma$ function and the dependence of all transcendental functions
on the matrices $U$ and $F$ is defined by their Taylor expansion.

We now wish to calculate the expectation value for vertex operators that correspond to different
closed string states, however this is a process that must be done with some care.  
To  calculate the emission of a closed string in the world-sheet picture
one generally  considers
 a disk emitting an asymptotic closed string state.  This is really a closed string 
cylinder 
diagram. The standard method is to use
 conformal invariance to map the closed string state to a point 
on
the disk, namely the origin, where a corresponding vertex operator is inserted.
On the other hand it has been cogently
argued that it is necessary to have an integrated vertex operator for closed string states to
properly couple
\cite{Craps:2001jp}, in particular that the graviton must be
produced by an integrated vertex operator to
couple correctly to the energy momentum tensor.  
The distinction between a fixed vertex operator and an integrated vertex
operator is moot in the conformally invariant case where the integration
will only produce a trivial volume factor, however in the case we consider
more care must be taken.  We wish to consider arbitrary locations of the
vertex operators on the string world sheet, and the natural measure to
impose is that of the conformal transformations which map the origin to
a point within the unit disk on the complex plane.

In other words we propose to allow the vertex
operator corresponding to the closed string state to be moved from the origin by a conformal
transformation that preserves the area of the unit disk, namely a PSL(2,R) transformation.
The method to accomplish this is to go to a new coordinate system
\beq
y = \frac{ a z + b }{b^* z + a^*},~~|a^2| - |b^2| = 1,
\eeq
and a vertex operator at the origin $y=0$ would correspond to an insertion of a vertex operator
at the point $z = \frac{-b}{a}$.  It is worth noting that in the case of conformal invariance,
that is when $U \rightarrow 0$ or $U \rightarrow \infty$ the greens function remains unchanged
in form, the $y$
dependence coming  from  the replacement $z \rightarrow z(y)$. Even in the case of finite $U$ the 
only change to the greens function is the addition of a term that is harmonic within the
unit disk.  The parameter of the integration over the position of the vertex operator
would be to the measure on PSL(2,R), giving an infinite factor in the conformally invariant
case \cite{Shatashvili:1993kk,Craps:2001jp,Liu:1988nz}.  
From this argument we have a definite prescription
for the calculation of vertex operator expectation values, which is to use the conformal 
transformation to modify the greens function, and calculate the expectation values of operators
at the origin with this modified greens function.

Now we will use this prescription to calculate the sigma model expectation values of some
operators, and we will start with the simplest, that of the closed string tachyon.
The vertex operator for the tachyon is $: e^{i p_\mu X^\mu(z(y))
} :$, and it is inserted at the point $y=0$.  
The normal ordering prescription for all such operators is 
that any divergent
pieces will be subtracted, but finite pieces will remain and by inspection we see that
the appropriately subtracted greens function is
\beqn
: {\cal G}^{\mu\nu}( z, z'): &=& G^{\mu\nu}(z,z') - g^{\mu\nu} \alpha' \ln \left| z - z'
\right| \nonumber \\
&=& 
\frac{\alpha'}{2}
\sum_{n=1}^\infty \left( \frac{g - 2 \pi \alpha' F - \frac{ \alpha'}{2}
\frac{U}{n} }{ g + 2 \pi \alpha' F + \frac{ \alpha'}{2} \frac{U}{n} } \right)^{\mu\nu}
\frac{(z \bar z')^n + (\bar z z')^n}{n}.
\label{subtractedgf}
\eeqn
Using (\ref{subtractedgf}) we can immediately see that
\beqn
\langle: e^{i p_\mu X^\mu(y=0) } : \rangle &=& 
 \left. Z e^{- \frac{1}{2} p_\mu p_\nu : {\cal G}^{\mu\nu}
(z(y), z'(y)) : } \right|_{y=0}
\nonumber \\
&=&
Z \exp \left( - \frac{\alpha'}{2} p_\mu p_\nu 
\sum_{n=1}^\infty \left( \frac{g - 2 \pi \alpha' F - \frac{ \alpha'}{2}
\frac{U}{n} }{ g + 2 \pi \alpha' F + \frac{ \alpha'}{2} \frac{U}{n} } \right)^{\mu\nu}
\frac{ 1}{n} \frac{ |b^{2n}|}{|a^{2n}|} \right).
\nonumber \\
&~& \label{tachyonEV}
\eeqn
Thus we find that the expectation value for the tachyon vertex operator is
\beqn
\int d^2a d^2b \delta( |a^2| - |b^2| -1 )
&~& \nonumber \\
\times \langle: e^{i p_\mu X^\mu(y(a,b)=0) } : \rangle &=& 
\int d^2a d^2b \delta( |a^2| - |b^2| -1 )
Z \nonumber \\
&~& 
\exp \left( - \frac{\alpha'}{2} p_\mu p_\nu 
\sum_{n=1}^\infty \left( \frac{g - 2 \pi \alpha' F - \frac{ \alpha'}{2}
\frac{U}{n} }{ g + 2 \pi \alpha' F + \frac{ \alpha'}{2} \frac{U}{n} } \right)^{\mu\nu}
\frac{ 1}{n} \frac{ |b^{2n}|}{|a^{2n}|}  \right).
\nonumber \\ &~& \label{inttachyonEV}
\eeqn

A similar analysis can also be performed for the massless closed string excitations.
In particular the graviton insertion at $y=0$ is given by 
\beqn 
\langle {\cal V}_h \rangle &=& \langle : - \frac{ 2}{\alpha'}
h_{\mu\nu} \dif X^\mu \bar \dif X^\nu e^{i p_\mu X^\mu(y=0)} : \rangle  
\eeqn  
where $h$ is a symmetric traceless tensor and the normalization follows the conventions
of \cite{Polchinski:1998rq}.
This can be analyzed by the same techniques as for the tachyon, noting that there
will be cross contractions between the exponential and the $X$-field prefactors.
Explicitly we obtain 
\beqn
\langle {\cal V}_h \rangle &=& 
- \frac{ 2}{\alpha'} Z h_{\mu\nu} \left( \dif \bar \dif' : {\cal G}^{\mu\nu}\left(z(y),z'(y)\right) :
+ \dif :{\cal G}^{\mu\alpha} \left(z(y), z'(y) \right): 
\right.
\nonumber \\ 
&~& \left. \times \bar \dif :{\cal G}^{\mu\beta} \left(z(y), z'(y) \right): (i 
p_\alpha) (i p_\beta)
\right)
e^{- \frac{1}{2} p_\mu p_\nu : {\cal G}^{\mu\nu} (z(y), z'(y) ) : }|_{y=0} 
\nonumber \\
&=& 
Z h_{\mu\nu}  \left( - 
\sum_{n=1}^\infty \left( \frac{g - 2 \pi \alpha' F - \frac{ \alpha'}{2}
\frac{U}{n} }{ g + 2 \pi \alpha' F + \frac{ \alpha'}{2} \frac{U}{n} } \right)^{\mu\nu}
n \frac{|b^{2 (n-1)}| }{|a^{2 (n-1)}| } \frac{1}{|a^2|^2}
\right. 
\nonumber \\
&~&  +
\frac{\alpha'}{2} 
\sum_{n=1}^\infty \left( \frac{g - 2 \pi \alpha' F - \frac{ \alpha'}{2}
\frac{U}{n} }{ g + 2 \pi \alpha' F + \frac{ \alpha'}{2} \frac{U}{n} } \right)^{\mu\alpha}
\frac{ |b^{2(n-1)} |}{ |a^{2(n-1)} |} \frac{-b}{|a^2|a}
\nonumber \\
&~& \times \left. 
\sum_{m=1}^\infty \left( \frac{g - 2 \pi \alpha' F - \frac{ \alpha'}{2}
\frac{U}{m} }{ g + 2 \pi \alpha' F + \frac{ \alpha'}{2} \frac{U}{m} } \right)^{\nu\beta}
\frac{ |b^{2(m-1)} |}{ |a^{2(m-1)} |} \frac{-b*}{|a^2|{a^*}} p_\alpha p_\beta
\right)
\nonumber \\
&~& 
\exp \left( - \frac{\alpha'}{2} p_\mu p_\nu 
\sum_{n=1}^\infty \left( \frac{g - 2 \pi \alpha' F - \frac{ \alpha'}{2}
\frac{U}{n} }{ g + 2 \pi \alpha' F + \frac{ \alpha'}{2} \frac{U}{n} } \right)^{\mu\nu}
\frac{ 1}{n} \frac{ |b^{2n}|}{|a^{2n}|}  \right).
\nonumber \\
&~& 
\label{gravitonEV}
\eeqn
Clearly a similar analysis can be performed for either the Kalb-Ramond field or the
dilaton, and the only change would be to replace $h_{\mu\nu}$ with the appropriate 
polarization tensor for either field.

Finally, we can perform the same kind of calculation for a more general closed string
state, and while the analysis below is not performed for a completely general state, 
it contains 
the germs of generality.
We consider a state which may be off shell in the sense that it not annihilated by 
the positive modes of the $\sigma$-model energy momentum tensor (the Virasoro
generators), may not satisfy the 
mass shell condition, and may
not be level matched. 
Our explicit choice is to consider the operator 
\beqn
\langle {\cal V}_A \rangle &=&
\langle : -i 
\left( \frac{2}{\alpha'} \right)^{3/2}
A_{\mu\nu\delta} \frac{\dif^a}{(a-1)!}  X^\mu \frac{\bar 
\dif^b}{(b-1)! }
X^\nu \frac{\bar \dif^c}{(c-1)!} X^\gamma e^{i p_\mu X^\mu } : \rangle
\eeqn
which is an arbitrary state involving three creation operators.
We find that
\beqn
\langle {\cal V}_A \rangle &=&
 Z A_{\mu\nu\delta} 
\left( \frac{2}{\alpha'} \right)^{3/2}
\left( \frac{\dif^a}{(a-1)!} \frac{\bar \dif'^b}{(b-1)!} 
: G^{\mu\nu}(z,z') : \frac{\bar \dif^c}{(c-1)!} :G^{\delta\alpha}(z,z'): p_\alpha
\right. \nonumber \\
&~& +  \frac{\dif^a}{(a-1)!} \frac{\bar \dif'^c}{(c-1)!}  
: G^{\mu\delta}(z,z') : \frac{\bar \dif^b}{(b-1)!} :G^{\nu\alpha}(z,z'): p_\alpha  
 \nonumber \\ &~&
\left. -  \frac{\dif^a}{(a-1)!} : G^{\mu\alpha}(z,z') :  \frac{\bar \dif^b}{(b-1)!}
:G^{\nu\beta}(z,z'): \frac{\bar \dif^c}{(c-1)!} :G^{\delta\gamma}(z,z'): 
p_\alpha p_\beta p_\gamma \right)
\nonumber \\ &~&
\left. \times e^{- \frac{1}{2} p_\mu p_\nu : {\cal G}^{\mu\nu} (z, z' ) : }
\right|_{y=0}.
\label{arbstateEV}
\eeqn
It is straightforward but not very instructive to take
 the derivatives 
acting on
the greens functions and evaluate the result at $y=0$.  This general state will allow
us to check the prescription for the boundary state presented in the next section.

\section{Boundary States}

We now perform the same kind of analysis from the point of view of the
boundary state formalism.  To this end, it is important to understand where the 
boundary state comes from.  By varying the action (\ref{action}) one obtains the 
equation
\beqn
\frac{1}{2\pi \alpha'} g_{\mu\nu} \dif_n X^\nu + F_{\mu\nu} \dif_t X^\nu
+ \frac{1}{4 \pi} U_{\mu\nu} X^\nu &=& 0
\label{boundcond}
\eeqn
on the boundary of the string world-sheet.  
As discussed in, for example, \cite{Lee:2001cs,DiVecchia:1999fx} 
the boundary state $|B\rangle$ 
is a state of closed string theory which obeys the boundary
condition (\ref{boundcond}) as an operator equation.
Using the following standard mode expansion for $X$ as a function of $z$
\beq
X^\mu(z,\bar z) = x^\mu + p^\mu \ln |z^2| + \sum_{m \neq 0} \frac{1}{m} \left(
\frac{\alpha_m^\mu}{z^m} + \frac{ \tilde \alpha_m^\mu}{\bar z^m} \right).
\eeq
we obtain as a boundary condition on the modes that
\beq
\left( g + 2\pi \alpha' F + \frac{\alpha'}{2} \frac{U}{n} \right)_{\mu\nu} \alpha^\mu_n
+ \left( g - 2\pi \alpha' F - \frac{\alpha'}{2} \frac{U}{n} \right)_{\mu\nu} \tilde 
\alpha^\mu_{-n} = 0.
\label{bosonicBC}
\eeq
We wish to make a state that obeys (\ref{bosonicBC}) as an operator equation, so that
operating with a state of positive index gives the appropriate negative index coefficient
to make
\beq
\left[\left( g + 2\pi \alpha' F + \frac{\alpha'}{2} \frac{U}{n} \right)_{\mu\nu} 
\alpha^\mu_n
+ \left( g - 2\pi \alpha' F - \frac{\alpha'}{2} \frac{U}{n} \right)_{\mu\nu} \tilde  
\alpha^\mu_{-n} \right] |B\rangle = 0,
\nonumber \\
\eeq
\beq
\left[ g_{\mu\nu} p^\mu - i \frac{\alpha'}{2} U_{\mu\nu} x^\mu \right] |B\rangle
 = 0. 
\label{conditiononP}
\eeq 
It is relatively easy to see that the state to satisfy this must be a coherent state
and is given by
\beq
|B\rangle = {\cal N} \prod_{n\geq1} \exp  \left( -
\left( \frac{g - 2 \pi \alpha' F - \frac{\alpha'}{2} \frac{U}{n} }{
 g + 2 \pi \alpha' F + \frac{\alpha'}{2} \frac{U}{n} } \right)_{\mu\nu}
\frac{\alpha^\mu_{-n} \tilde \alpha^\nu_{-n} }{n} \right) 
\exp \left(- \frac{\alpha'}{4} x^\mu U_{\mu\nu} x^\nu \right) | 0 \rangle
\eeq
where ${\cal N}$ is an as yet undetermined normalization constant.

It is interesting to examine how this boundary state transforms under the 
action of the residual conformal symmetry of the disk, namely under PSL(2,R)
transformations.  In the two conformally invariant cases ($U=0$ and $U=\infty$)
this is a good symmetry of the action, but we expect there to be interpolation
as the flow from $U=0$ to $U=\infty$ takes us from Dirichlet to Neumann 
boundary conditions.   As mentioned previously the 
action of  PSL(2,R) on the complex coordinates of the
disk is to perform the mapping
\beq
z \rightarrow w(z) = \frac{a z + b}{b^* z + a^*}
\eeq
where $a$ and $b$ satisfy the relation
\beq
|a^2| - |b^2| = 1.
\eeq
This transformation maps the interior of the unit disk to itself, the exterior
to the exterior and the boundary to the boundary.
Moreover, this transformation of the coordinates induces a 
mapping which intermixes the oscillator modes.
To see this consider the definition of the oscillator modes
\beq
\alpha_m^\mu = \sqrt{ \frac{2}{\alpha'} }
\oint \frac{dz}{2\pi } z^m \dif X^\mu(z)
\eeq
where the contour is the boundary of the unit disk, and the mode expansion of
$X$ is 
\beq
\dif X^\mu(z) = -i \sqrt{ \frac{\alpha'}{2} } \sum_m \frac{\alpha_m^\mu}{z^{m+1} }.  
\label{modeexp}
\eeq
Now, using the fact that $X$ is a scalar, or equivalently the fact that 
$\dif X$ is a (1,0) tensor, we see that
\beq
\alpha_m^\mu = \oint \frac{dz}{2\pi i} z^m \dif_w X^\mu(w) \frac{dw}{dz}
.
\eeq
Now, using the fact that a mode expansion for $X$ exists in terms of $w$ with 
coefficients $\alpha'_m$ in exactly the same way as (\ref{modeexp}), we see that
\beq
\alpha_m^\mu = M_{mn} \alpha_n^{'\mu} 
\eeq
where 
\beq
M_{mn} = \oint \frac{dz}{2\pi i} z^m \frac{ (b^* z + a^*)^{n-1} }{ 
(a z + b)^{n+1} }.  
\label{defofM}
\eeq
Upon evaluation it 
becomes clear that this matrix has a block diagonal form, which is
 due to the fact that there
are no poles inside (outside) the integration region when $n<0<m$ ($m<0<n$).  
The consequence of this is that under PSL(2,R) transformations
the creation operators transform into creation operators and the annihilation
operators likewise transform into annihilation operators.
Furthermore, upon rescaling ${\cal M}_{mp} \rightarrow \sqrt{\frac{p}{m} } M_{mp}$ 
so that the oscillators are normalized as creation
and annihilation operators it is easy to show that for both
positive and negative $p$ and $m$ 
\beq
{\cal M}^{-1}_{mp} = \oint \sqrt{ \frac{m}{p} } 
\frac{d \bar z }{-2 \pi i} \bar z^p \frac{ 
(a + b \bar z)^{m-1} }{
(b^* + a^* \bar z )^{m+1} } 
= {\cal M}_{mp}^{*T} = {\cal M}_{mp}^\dagger.
\eeq
This is simply a statement of the fact that $M$ preserves the inner product on
the space of operators.  It would be inappropriate for $M$ itself to be hermitian
because the commutation relation between the modes is 
\beq 
\left[ \alpha^\mu_n, \alpha^\nu_m \right] = n \delta_{n-m} g^{\mu\nu}.
\nonumber 
\eeq 
The general expression for $M$ when both indices are positive can easily be found
by explicit contour integration and is
\beq
M_{mp} = \sqrt{ \frac{p}{m} } \sum_{k=0}^p 
\left( \matrix{ m \cr k } \right) 
\left( \matrix{ p-1 \cr p-k } \right)  
(-1)^{m-k} b^m |b^2|^{-k} a^{-p},
\eeq
where the binomial coefficients $\left( \matrix{ a \cr b } \right)$ are understood
to vanish in all cases where $b>a$.
It will be important to note that there is a non-zero overlap with the the zero mode
which will be important when we make a correspondence between the sigma model calculation
and the boundary state. 

To accommodate the coordinate transformation under PSL(2,R) the boundary state becomes
\beqn
|B_{a,b} \rangle &=& {\cal N} \exp \left( \sum_{n=1, j,k=-\infty}^\infty 
\alpha^{\mu}_{-k} M_{-n-k}(a,b) 
\left( \frac{g - 2 \pi \alpha' F - \frac{\alpha'}{2} \frac{U}{n} }{
 g + 2 \pi \alpha' F + \frac{\alpha}{2} \frac{U}{n} } \right)_{\mu\nu}
\right. 
\nonumber \\ &~& \left. \frac{1}{n} M^*_{-n-j}(a,b)
 \tilde \alpha^{'\nu}_{-j}  \right)
\exp \left(- \frac{\alpha'}{4} x^\mu U_{\mu\nu} x^\nu \right)
 | 0 \rangle
\label{transfboundstate}
\eeqn
and in this equation and all following ones we drop the ' associated with the
transformation  for notational simplicity.  We 
conjecture that the proper definition of the boundary state to give
the correct overlap with all closed string states is
\beq
|B\rangle = \int d^2 a d^2 b \delta(|a^2| - |b^2| -1) |B_{a,b} \rangle. 
\eeq
This is just the boundary state 
(\ref{transfboundstate}) integrated over the Haar measure of  PSL(2,R).
We will now verify that this gives the correct overlap with the tachyon and massless
states by comparing with the $\sigma$-model calculations of the previous section.

The computation for the tachyon is easy.  To fix the normalization it suffices
to take the inner product
\beqn
\langle 0 | B \rangle &=& {\cal N}  \int d^2 a d^2 b \delta(|a^2| - |b^2| -1)
\exp \left(- \frac{\alpha'}{2} x^\mu U_{\mu\nu} x^\nu \right) 
\nonumber \\
&~& \exp \left( - \sum_{n=1}^\infty \frac{\alpha'}{2} p_\mu M_{-n0} \frac{1}{n}
\left( \frac{g - 2 \pi \alpha' F - \frac{\alpha'}{2} \frac{U}{n} }{
 g + 2 \pi \alpha' F + \frac{\alpha}{2} \frac{U}{n} } \right)_{\mu\nu}
M^*_{-n0} p_\nu \right)
\label{tachyonBScalc}
\eeqn
and we have used the relations $\alpha_0^\mu = \sqrt{\frac{\alpha'}{2} } p^\mu$ and
$\left[ x^\mu, p^\nu \right] = i g^{\mu\nu}$.
We can insert the explicit form  $M_{-n0} = \left( \frac{-b^*}{a^*} \right)^n$ to
(\ref{tachyonBScalc}) and find 
\beqn
\langle 0 | B \rangle &=& {\cal N}  \int d^2 a d^2 b \delta(|a^2| - |b^2| -1)
\exp \left(- \frac{\alpha'}{2} x^\mu U_{\mu\nu} x^\nu \right)
\nonumber \\
&~& \exp \left( - \sum_{n=1}^\infty \frac{\alpha'}{2} p_\mu \frac{1}{n}
\left( \frac{g - 2 \pi \alpha' F - \frac{\alpha'}{2} \frac{U}{n} }{
 g + 2 \pi \alpha' F + \frac{\alpha}{2} \frac{U}{n} } \right)_{\mu\nu}  p_\nu
\frac{ |b^{2n}| }{|a^{2n} |} \right).
\eeqn 
By comparing this to (\ref{inttachyonEV}) we can unambiguously fix the normalization as
\beq
{\cal N} = Z
.
\eeq 

We now perform an analogous calculation for the massless states
which will provide a non-trivial
check of this normalization scheme.
For an arbitrary massless state with polarization tensor $P_{\mu\nu}$
\beqn 
| P_{\mu\nu} \rangle &=&  P_{\mu\nu} \alpha^\mu_{-1}
\tilde \alpha^\nu_{-1} | 0 \rangle
\label{masslessdefn}
\eeqn
 the overlap
to be calculated is 
\beqn
\langle   P_{\mu\nu} | B \rangle 
&=& 
{\cal N}  
\int d^2 a d^2 b \delta(|a^2| - |b^2| -1)
\exp \left(- \frac{\alpha'}{2} x^\mu U_{\mu\nu} x^\nu \right)
\nonumber \\
&~& \exp 
\left( - \sum_{n=1}^\infty \frac{\alpha'}{2} p_\mu M_{-n0} \frac{1}{n}  
\left( \frac{g - 2 \pi \alpha' F - \frac{\alpha'}{2} \frac{U}{n} 
}{ g + 2 \pi \alpha' F + \frac{\alpha}{2} \frac{U}{n} } \right)_{\mu\nu}
M^*_{-n0} p_\nu \right)
\nonumber \\
&~& P_{\mu\nu}  \left[ - \sum_{n=1}^\infty   M_{-n-1}
\frac{1}{n}
\left( \frac{g - 2 \pi \alpha' F - \frac{\alpha'}{2} \frac{U}{n}
 }{ g + 2 \pi \alpha' F + \frac{\alpha}{2} \frac{U}{n} } \right)^{\mu\nu}
M^*_{-n-1}   \right.
\nonumber \\
&~& +\frac{\alpha' }{2} \sum_{n=1}^\infty   M_{-n-1}
\frac{1}{n}
\left( \frac{g - 2 \pi \alpha' F - \frac{\alpha'}{2} \frac{U}{n} }{
 g + 2 \pi \alpha' F + \frac{\alpha}{2} \frac{U}{n} } \right)^{\mu\alpha}
M^*_{-n0} p_\alpha
\nonumber \\
&~& \times \left. \sum_{m=1}^\infty   M_{-m0}
\frac{1}{m} \left( \frac{g - 2 \pi \alpha' F - \frac{\alpha'}{2} \frac{U}{m} }{
 g + 2 \pi \alpha' F + \frac{\alpha}{2} \frac{U}{m} } \right)^{\beta\nu} M^*_{-n-1}
p_\beta
\right]
\nonumber \\
&=&
{\cal N}  \int d^2 a d^2 b \delta(|a^2| - |b^2| -1)
\exp \left(- \frac{\alpha'}{2} x^\mu U_{\mu\nu} x^\nu \right)
\nonumber \\
&~& \exp \left( - \sum_{n=1}^\infty \frac{\alpha'}{2} p_\mu \frac{1}{n}
\left( \frac{g - 2 \pi \alpha' F - \frac{\alpha'}{2} \frac{U}{n} }{
 g + 2 \pi \alpha' F + \frac{\alpha}{2} \frac{U}{n} } \right)_{\mu\nu}  p_\nu
\frac{ |b^{2n}| }{|a^{2n} |} \right)
\nonumber \\
&~&
 P_{\mu\nu}  \left[ - \sum_{n=1}^\infty   
\left( \frac{g - 2 \pi \alpha' F - \frac{\alpha'}{2} \frac{U}{n}
 }{ g + 2 \pi \alpha' F + \frac{\alpha}{2} \frac{U}{n} } \right)^{\mu\nu}
n \frac{ |b^{2(n-1)}| }{ |a^{2(n-1) }| } \frac{1}{|a^2|^2}   \right.
\nonumber \\
&~& +\frac{\alpha' }{2} \sum_{n=1}^\infty  
\left( \frac{g - 2 \pi \alpha' F - \frac{\alpha'}{2} \frac{U}{n} }{
 g + 2 \pi \alpha' F + \frac{\alpha}{2} \frac{U}{n} } \right)^{\mu\alpha}
\frac{ |b^{2(n-1)}| }{ |a^{2(n-1) }| } \frac{ -b^* }{ |a^2| a^* } p_\alpha
\nonumber \\
&~& 
\times \left. \sum_{m=1}^\infty  
 \left( \frac{g - 2 \pi \alpha' F - \frac{\alpha'}{2} \frac{U}{m} }{
 g + 2 \pi \alpha' F + \frac{\alpha}{2} \frac{U}{m} } \right)^{\beta\nu} 
\frac{ |b^{2(m-1)}| }{ |a^{2(m-1) }| } \frac{ -b}{ |a^2| a}
p_\beta
\right]
\label{gravitonBScalc}
\eeqn
Now, we compare 
(\ref{gravitonBScalc}) with (\ref{gravitonEV}) to again
find that the normalization is fixed by
\beqn
 {\cal N} = Z
.
\nonumber
\eeqn

We can also use the boundary state to compute the overlap with the more general
state that we considered in the sigma model calculations.
Explicitly the overlap between the boundary state and state $A$ defined by
\beqn 
|A_{\mu\nu\delta} \rangle &=&  
 A_{\mu\nu\delta} \alpha_{-a}^\mu \tilde \alpha_{-b}
^\nu \tilde \alpha_{-c}^\delta |0\rangle
\eeqn
 is given as
\beqn
\langle  A_{\mu\nu\delta}
| B \rangle 
&=&
{\cal N}
\int d^2 a d^2 b \delta(|a^2| - |b^2| -1)
\exp \left(- \frac{\alpha'}{2} x^\mu U_{\mu\nu} x^\nu \right)
\nonumber \\
&~& \exp 
\left( - \sum_{n=1}^\infty \frac{\alpha'}{2} p_\mu M_{-n0} \frac{1}{n}
\left( \frac{g - 2 \pi \alpha' F - \frac{\alpha'}{2} \frac{U}{n}
}{ g + 2 \pi \alpha' F + \frac{\alpha}{2} \frac{U}{n} } \right)_{\mu\nu}
M^*_{-n0} p_\nu \right)
\nonumber \\
&~&
A_{\mu\nu\delta} \sqrt{\frac{\alpha'}{2}}
 \left[ \sum_n ab M_{-n -a} 
\Lambda^{\mu\nu}(n) 
M^*_{-n-b}
\sum_m c M_{-m 0} \Lambda^{\alpha \delta}(m) M^*_{-m-c} p_\alpha \right.
\nonumber \\
&~&  + \sum_n ac M_{-n -a} \Lambda^{\mu\delta}(n) M^*_{-n-c}
\sum_m b M_{-m 0} \Lambda^{\alpha \nu}(m) M^*_{-m-b} p_\alpha
\nonumber \\ &~&
 - \frac{\alpha'}{2} \sum_n aM_{-n -a} \Lambda^{\mu\alpha}(n) M^*_{-n0}
\sum_m bM_{-m 0} \Lambda^{\beta \nu}(m) M^*_{-m-b}
\nonumber \\ &~&
\left. \times \sum_l c M_{-l 0} \Lambda^{\gamma \delta}(l) M^*_{-l-c} p_\alpha p_\beta p_\gamma
\right],
\label{arbstateBScalc}
\eeqn
and we have introduced the notational simplification
\beq
\frac{1}{m} \left( \frac{g - 2 \pi \alpha' F - \frac{\alpha'}{2} \frac{U}{m} }{   
 g + 2 \pi \alpha' F + \frac{\alpha}{2} \frac{U}{m} } \right)^{\mu\nu} = \Lambda^{\mu\nu}(m).
\eeq
We would like to compare (\ref{arbstateEV}) to (\ref{arbstateBScalc}) in the same manner that 
we have for the tachyon and the massless states, and to do so requires a simple calculation.
It can be shown that 
\beq 
\frac{1}{(k-1)!} \dif^k z^d(y) |_{y=0} = k M^*_{-d-k}.
\eeq
This shows that the result obtained from differentiating the greens function, as deformed in
the sigma model, gives the same result as the overlap of a closed string oscillator mode with
the boundary state.  Finally, again direct comparison of 
(\ref{arbstateEV}) to (\ref{arbstateBScalc}) gives ${\cal N} = Z$ again.
This shows that our 
boundary state $|B\rangle$ gives the
correct overlap with any closed string state, either on shell or off shell, and has fixed the
normalization to be equal to the partition function.

\section{One loop Boundary States}

Now that we have fixed the normalization of the boundary states, and provided a consistent
prescription for the action of the arbitrary PSL(2,R) transformation, we turn to the
more intricate subject of the calculation of the overlap of two such states.
This calculation can be thought of as a tree level exchange of closed strings 
between two D-branes with arbitrary field content.  The naive expectation is the following, that
the resulting expression will be the contribution of all the possible on and off shell one
particle states that the boundary state can emit, weighted by the closed string propagator.
It is interesting to note that, depending on $U$,
 not all possible physical closed string excitations
are produced by the boundary state, and that in general 
 states that do not satisfy the physical
state condition are created.

Explicitly the thing we wish to calculate is
the open string one loop correction to the partition function.  The disk level correction was
given in (\ref{partitionfunction}) and the one loop correction is given in the sigma model 
calculation by using the propagator on an annulus world-sheet, however in the boundary
state representation the calculation is
\beqn
Z_{One~loop} &=& \int d^2a d^2b \delta( |a^2| - |b^2| -1) d^2a' d^2b' \delta (|a^{'2}| - |b^{
'2}| -1 ) \times \nonumber \\
&~& {\cal N}^2 \langle B_{a',b'} | \frac{1}{L_0 + \tilde L_0 -2} | B_{a,b} \rangle.
\eeqn
In the above, $| B_{a,b} \rangle$ is as given in 
(\ref{transfboundstate}).  Note that this formulation will explicitly give factorization
of the amplitude in the closed string channel.

The calculation of $Z_{One~loop}$ is a straightforward, albeit tedious exercise.  
Using an integral representation for the propagator we find
\beqn
Z_{One~loop} &=& \int d^2a d^2b \delta( |a^2| - |b^2| -1) d^2a' d^2b' \delta (|a^{'2}| - |b^{
'2}| -1 ) \int_0^\infty dt \times \nonumber \\
&~& {\cal N}^2 \langle B_{a',b'} | e^{-t ( L_0 + \tilde L_0 -2 )} | B_{a,b} \rangle
.
\eeqn
Now, suppressing for the moment the integrals, it is necessary to calculate the inner
product itself, that is 
\beqn
 \langle B_{a',b'} | e^{-t ( L_0 + \tilde L_0 -2 )} | B_{a,b} \rangle
&=& \langle - i \frac{\alpha'}{2} U^{\mu\nu} x_\nu |
 \exp \left( \sum_{n=1, j,k=-\infty}^\infty
\alpha^{\mu}_{k} {M_{-n-k}^{{(1)}*}}
\Lambda_{\mu\nu}(n)
 M^{(1)}_{-nj}
 \tilde \alpha^{\nu}_{j}  \right)
\nonumber \\
&~& \exp \left( -t \sum_{n\geq 1} \left(\alpha^\mu_{-n} \alpha_{n\mu} + \tilde \alpha^\mu_{-n}
\tilde \alpha_{n\mu}\right) - \frac{t \alpha'}{4} p^\mu p_\mu \right)
\nonumber \\
&~&
\exp \left( \sum_{n=1, j,k=-\infty}^\infty
\alpha^{\mu}_{-k} {M_{-n-k}^{(2)}}  
\Lambda_{\mu\nu}(n)
 {M^{{(2)}*}_{-nj}}     
 \tilde \alpha^{\nu}_{-j}  \right)
|- i \frac{\alpha'}{2}  U^{\mu\nu} x_\nu
  \rangle.
\nonumber \\
\eeqn
In this, we have denoted the dependence on a particular SL(2,R) transform by a superscript on the
appropriate matrix, and abbreviated the Gaussian term in $x$ acting on the Fock space vacuum.
The above expression can be simplified considerably by using  the relation
\beqn
e^A e^B &=& e^B \left( \prod_{n=1}^\infty e^{\frac{1}{n!}[A, \ldots , [A,B]]}\right) e^A
\eeqn
which holds when $[A,B],~[A,[A,B]]$ and all similarly nested commutators commute with each other
and with $B$, but do not commute with $A$.
This then gives
\beqn
 \langle B_{a',b'} | e^{-t ( L_0 + \tilde L_0 -2 )} | B_{a,b} \rangle
&=& \langle - i \frac{\alpha'}{2} U^{\mu\nu} x_\nu |
 \exp \left( \sum_{n=1, j,k=-\infty}^\infty
\alpha^{\mu}_{k} {M_{-n-k}^{{(1)}*}}
\Lambda_{\mu\nu}(n)
 M^{(1)}_{-nj}
 \tilde \alpha^{\nu}_{j}  \right)
\nonumber \\
&~&
 \exp \left( - \frac{t \alpha'}{4} p^\mu p_\mu \right)
\exp \left( \sum_{n=1, j,k=-\infty}^\infty
\alpha^{\mu}_{-k} e^{-tk} {M_{-n-k}^{(2)}}
\Lambda_{\mu\nu}(n)
\right. \nonumber \\
&~& \left. {M^{{(2)}*}_{-nj}}
 \tilde \alpha^{\nu}_{-j} e^{-tj}  \right)
|- i \frac{\alpha'}{2}  U^{\mu\nu} x_\nu
  \rangle.
\nonumber \\
\eeqn
From here it is a straightforward exercise in combinatorics and commutation, 
very reminiscent of proofs of Wick's 
theorem, to obtain
\beqn
 \langle B_{a',b'} | e^{-t ( L_0 + \tilde L_0 -2 )} | B_{a,b} \rangle
&=&
 e^{2t} 
\exp  \sum_k \frac{1}{k} \delta_{\mu}^{\nu} \delta_{mn} \left(
\left[ M_{-m-a}^{(2)} a e^{-ta} M^{{(1)}*}_{-k-a} \Lambda^\mu_\alpha(k)
\right. \right. \nonumber \\
&~& \left. \left.  M^{(1)}_{-k-b} b e^{-tb} M^{{(2)}*}_{-n-b}
\Lambda^\alpha_\nu(n) 
\right]^k 
\right)^{\mu\nu}_{mn} F(x)  
\label{annulusBSamp}
\eeqn
with matrix multiplication implied within the sum for both the Lorentz and oscillator indices, and
$F(x)$ a Gaussian that depends on $x, U, F$ and the transformation parameters.
It can be verified that 
\beqn
F(x) &=& \exp p^\mu p^\nu \left( -\frac{t \alpha'}{4} g_{\mu\nu}  + 
\right. 
M^{{(1)}*}_{-k0} \Lambda^\mu_\alpha(k) 
   M^{(1)}_{-k-b} \times \nonumber \\ &~& 
\frac{1}{1 - 
b e^{-tb} M^{{(2)}*}_{-n-b} \Lambda^\alpha_\beta(n)  M_{-n-a}^{(2)} a e^{-ta}
M^{{(1)}*}_{-p-a} \Lambda^\beta_\gamma(p)  M^{(1)}_{-p-q} } q e^{-tq} 
M^{{(2)}*}_{-r-q} 
 \Lambda^\gamma_\nu(q)  M_{-q0}^{(2)} + 
\nonumber \\ &~&
M^{{(1)}*}_{-k0} \Lambda^\mu_\alpha(k)  M^{(1)}_{-k-b} \frac{1}{1 -
b e^{-tb} M^{{(2)}*}_{-n-b} \Lambda^\alpha_\beta(n)  M_{-n-a}^{(2)} a e^{-ta}
M^{{(1)}*}_{-p-a} \Lambda^\beta_\nu(p)  M^{(1)}_{-p-0} }
+ 
\nonumber \\ &~& \left. \frac{1}{1 -
 M^{{(2)}*}_{-n0} \Lambda^\mu_\beta(n)  M_{-n-a}^{(2)} a e^{-ta}
M^{{(1)}*}_{-p-a} \Lambda^\beta_\gamma(p)  M^{(1)}_{-p-q} } q e^{-tq} 
M^{{(2)}*}_{-r-q}
 \Lambda^\gamma_\nu(q)  M_{-q0}^{(2)}
\right),
\eeqn
with $p^\mu = - \frac{i \alpha'}{2} U^\mu_\nu x^\nu$ as in
\ref{conditiononP} .  It should be noted that this seemingly 
complicated expression is nothing but a Gaussian in $x$, and so just has the effect of localizing
the interaction between two D-branes.
The final expression for $Z_{One~loop}$ is obtained by integrating (\ref{annulusBSamp}) over the
elements of the conformal transformations noted 
above. 

It is instructive to examine the form of this.  
First, note that ignoring the $x$ dependence we can  express as
\beqn
Z_{One~loop} &=& \int d^2a d^2b \delta( |a^2| - |b^2| -1) d^2a' d^2b' \delta (|a^{'2}| - |b^{
'2}| -1 ) \times \nonumber \\
&~& {\cal N}^2
 e^{2t} \frac{1}{\det \left( 1- 
b e^{-tb} M^{{(2)}*}_{-n-b} \Lambda^\alpha_\beta(n)  M_{-n-a}^{(2)} a e^{-ta}
M^{{(1)}*}_{-p-a} \Lambda^\beta_\gamma(p)  M^{(1)}_{-p-q}
\right)  }
\nonumber \\
\eeqn
In the cases of $U \rightarrow 0$ and $U \rightarrow \infty$ 
the matrices $M^*\Lambda M$ revert to a particularly simple form.
We have
\beqn \left.
\sum_{n \geq 1}
M^{*}_{-n-b} \Lambda^{\mu\nu}(n)  M_{-n-a} \right|_{(U \rightarrow 0)} &=& 
\sum_{n \geq 1} M^*_{-n-b}
\frac{1}{n} \left( \frac{g - 2 \pi \alpha' F  }{
 g + 2 \pi \alpha' F  } \right)^{\mu\nu} M_{-n -a}
\nonumber \\
&=& \left( \frac{g - 2 \pi \alpha' F  }{  
 g + 2 \pi \alpha' F  } \right)^{\mu\nu} \frac{1}{b} \delta_{ab}
\eeqn
and also 
\beqn \left.
 \sum_{n \geq 1}
M^{*}_{-n-b} \Lambda^{\mu\nu}(n)  M_{-n-a} \right| _{(U \rightarrow \infty)}
&=& -g^{\mu\nu} \frac{1}{b} \delta_{ab}
\eeqn
which follow from the definition of $M$.
We can see that in the case of $U = 0$ that the boundary state for 
a background gauge field is recovered, and in the case $U = \infty$
the boundary state for a localized object, a D-brane, is recovered.
In both the cases the integral over the volume of PSL(2,R) becomes
a trivial prefactor, as expected.
It is also possible to say something about the more general case.   
The sum $
M_{-n-a}^{(2)} a e^{-ta}
M^{{(1)}*}_{-q-a}
$ can be taken over $a$ and we find the result is a transformation
in SL(2,C), of which PSL(2,R) is a subgroup.
Explicitly
\beqn
\sum_{a \geq 1} M_{-n-a}^2 a e^{-ta}
M^{{(1)}*}_{-q-a}
&=&
\oint \frac{d z}{ 2 \pi i} \frac{1}{z^n} 
\frac{ \left[ z (a_1^* e^{-t/2} a_2 - b_1 e^{t/2} b_2^*) + (a^*_1 e^{-t/2} b_2 
- b_1 e^{t/2} a_2^* ) \right]^{q-1} }{
\left[ z (a_1 e^{t/2} b_2^* - b_1^* e^{-t/2} a_2 ) + (a_1 e^{t/2} a_2^* -
b_1^* e^{-t/2} b_2) \right]^{q+1}  }.
\nonumber \\
\eeqn
The fact that the integral over conformal factors becomes an integral over a transformation
in a larger group is appropriate.  The integral over $t$ stands in the place
of the integral over the Teichmuller parameter of the annulus, and the other
degrees of freedom correspond to reparametrizations of the two ends of the annulus.

It is also interesting to examine how these expressions for 
$Z_{One~loop}$ vary with $U$ around the two fixed points.
In particular, ignoring the linear terms in $U$ in the normalization, which
can be seen (\ref{partitionfunction}) to be divergent, the expression
for $Z_{One~loop}$ near $U=0$ is
\beqn
Z_{One~loop} &=& Z_{One~loop}(U=0) + 
Tr\left( U \frac{\partial}{\partial U} Z_{One~loop}(U=0) \right)
 + \ldots
\eeqn
Immediately upon differentiation we see that the linear term will be given by
\beqn
Tr\left( U \frac{\partial}{\partial U} Z_{One~loop}(U=0) \right)
 &=& \int d^2a d^2b \delta( |a^2| - |b^2| -1) {\cal V}_{PSL(2,R)}
{\cal N}^2
 e^{2t} 
\nonumber \\ &~&
\frac{1}{\det \left( 1- 
e^{-2tb} \left( \frac{ g - 2 \pi \alpha' F}{g+2 \pi \alpha' F} \right)^2
\right)  }
\nonumber \\
&~& \times Tr\left( \frac{ (g- 2\pi \alpha' F )/(g + 2\pi \alpha' F)}{
(g + 2\pi \alpha' F)^2 - (g- 2\pi \alpha' F)^2 } U \right) 
\nonumber \\ &~& \times \frac{-4 b e^{-2b} }{1 - e^{-2 b} }
M^*_{-n-b} \frac{1}{n^2} M_{-n-b}.
\eeqn
The factor of $\frac{e^{-2b}}{ 1- e^{-2b} }$ comes from the fact that all the other
$\Lambda$ terms become trivial because we have evaluated them at $U=0$
which was noted to be conformally invariant, and from summing the terms $e^{-b}$ which
stand between these.  Likewise note that the factor $1/n^2$ instead of $1/n$ between $M^*$ and $M$ comes 
from the fact that $U$ enters always as $U/n$. 
Also, one of the integrals over the PSL(2,R) groups becomes trivial, and relabeling gives the 
factor ${\cal V}_{PSL(2,R)}$ and only one integral.
Now, to calculate explicitly 
\beqn 
\sum_{n\geq 1} M^*_{-n-b} \frac{1}{n^2} M_{-n-b}
&=&
\sum_{n \geq 1} \oint \frac{dz }{2 \pi i} \frac{ d \bar z}{ -2 \pi i} 
\frac{1}{n^2} \frac{1}{ z^n \bar z^n }
\nonumber \\ &~& 
\frac{ (a^* \bar z + b^*)^{b-1} }{(b \bar z + a )^{b+1} }
\frac{ (a  z + b)^{b-1} }{(b^* z + a^* )^{b+1} }
\nonumber \\
&~&
\frac{1}{n!^2} \partial_z^{n-1} \partial_{\bar z}^{n-1} 
\left.\frac{ (a^* \bar z + b^*)^{b-1} }{(b \bar z + a )^{b+1} }
\frac{ (a  z + b)^{b-1} }{(b^* z + a^* )^{b+1} }
\right|_{z,\bar z = 0}
\eeqn
This can be calculated explicitly, and when we include the
 integration over PSL(2,R) we find that
it becomes
\beqn
 \int d^2a d^2b \delta( |a^2| - |b^2| -1) &~&
\nonumber \\
\times \sum_{n\geq 1} M^*_{-n-b} \frac{1}{n^2} M_{-n-b}
&=&
 \int d^2a d^2b \delta( |a^2| - |b^2| -1) 
\sum_{n\geq 1} \sum_{q=0}^{min( n-1, b-1)} \frac{1}{(nb)^2} 
\nonumber \\ &~&
\left( \frac{ b+n-q-1 !}{q! n-q-1 ! b-q-1!} 
\right)^2 \left( \frac{ |b^2 |}{|a^2| } \right)^{b+n-2q-2} \frac{1}{|a^2|^2} 
\nonumber \\
\eeqn
and we have used the fact that upon integration over the phase of $a$ and $b$ we will
have  orthogonality 
in the sum.
All in all the 
contribution is 
\beqn
Tr\left( U \frac{\partial}{\partial U} Z_{One~loop} \right)_{U=0}
 &=& 
\int d^2a d^2b \delta( |a^2| - |b^2| -1) {\cal V}_{PSL(2,R)}
{\cal N}^2
 e^{2t} 
\nonumber \\
&~& \times \frac{1}{\det \left( 1- 
e^{-2tb} \left( \frac{ g - 2 \pi \alpha' F}{g+2 \pi \alpha' F} \right)^2
\right)  }
\nonumber \\
&~& 
\times Tr\left( \frac{ (g- 2\pi \alpha' F )/(g + 2\pi \alpha' F)}{
(g + 2\pi \alpha' F)^2 - (g- 2\pi \alpha' F)^2 } U \right) 
\sum_{n,b \geq 1}
\frac{-4 b e^{-2b} }{1 - e^{-2 b} }
\nonumber \\
&~&
\times  \sum_{q=0}^{min( n-1, b-1)} \frac{1}{(nb)^2} \left( \frac{ b+n-q-1 !}{q! n-q-1 ! b-q-1!} 
\right)^2 
\nonumber \\ 
&~& \left( \frac{ |b^2 |}{|a^2| } \right)^{b+n-2q-2} \frac{1}{|a^2|^2}. 
\label{varabout0}
\eeqn

A similar calculation can be done around the d-brane ($U \rightarrow \infty$) and we find
that
\beqn
Tr\left( \frac{1}{U} \frac{\partial}{\partial (1/U)} Z_{One~loop} \right)_{\frac{1}{U} = 0} 
&=&
\int d^2a d^2b \delta( |a^2| - |b^2| -1) {\cal V}_{PSL(2,R)}
 \frac{{\cal N}^2
 e^{2t}}{\det \left( 1- 
e^{-2tb} 
\right)  }
\nonumber \\ &~&
\times 4 Tr\left( 
\frac{1}{ U} \right) 
\sum_{n,b \geq 1}
\frac{b e^{-2b} }{1 - e^{-2 b} }
M^*_{-n-b} M_{-n-b}.
\label{varaboutbig}
\eeqn
Because the natural coefficient for $\frac{1}{U}$ is $n$ the $n$ dependence between the
matrices $M$ is suppressed.
Explicit evaluations show that $
M^*_{-n-a} M_{-n-b}$ has zero entries on diagonal, so this variation vanishes about the 
d-brane.
This comparison between
(\ref{varabout0}) and (\ref{varaboutbig})
shows that the case of Neumann boundary conditions, (corresponding 
to $U \rightarrow
0$) is unstable with respect to variations of the tachyon condensate
since the linear variation does not vanish, but that Dirichlet boundary
conditions, obtained as $U \rightarrow \infty$ are stable.  This illustrates the well known
phenomenon of tachyon condensation and gives a mechanism to see explicitly how the 
open string tachyon has been removed from the excitations of the condensed state.

\section{Conclusions}

In this note we have presented a generalization of the boundary state formalism that
allows us to calculate the overlap of the boundary state with arbitrary on and off shell closed
string states.  We have shown that it exactly reproduces the calculations that would
be done in a sigma model for an appropriate vertex operator, verifying the conjecture of
\cite{Akhmedov:2001yh}.  This 
generalization gives
a prescription for the calculation of the boundary state annulus amplitude which reproduces
the expected modular transformation structure and explicitly factorizes in the closed string
channel.

\section{Acknowledgments}

The authors are grateful to Emil Akhmedov and Taejin Lee for many helpful 
conversations on this work.
This work supported in part by the Canadian 
Natural Science and Engineering Research Council.

\bibliography{tc1ln}

\begin{thebibliography}{10}

\bibitem{Witten:1992qy}
Edward Witten.
\newblock On background independent open string field theory.
\newblock {\em Phys. Rev.}, D46:5467--5473, 1992.

\bibitem{Witten:1993cr}
Edward Witten.
\newblock Some computations in background independent off-shell string theory.
\newblock {\em Phys. Rev.}, D47:3405--3410, 1993.

\bibitem{Witten:1993ed}
Edward Witten.
\newblock Quantum background independence in string theory.
\newblock 1993.

\bibitem{Shatashvili:1993kk}
Samson~L. Shatashvili.
\newblock Comment on the background independent open string theory.
\newblock {\em Phys. Lett.}, B311:83--86, 1993.

\bibitem{Shatashvili:1993ps}
Samson~L. Shatashvili.
\newblock On the problems with background independence in string theory.
\newblock 1993.

\bibitem{Gerasimov:2000zp}
Anton~A. Gerasimov and Samson~L. Shatashvili.
\newblock On exact tachyon potential in open string field theory.
\newblock {\em JHEP}, 10:034, 2000.

\bibitem{Gerasimov:2000ga}
Anton~A. Gerasimov and Samson~L. Shatashvili.
\newblock Stringy higgs mechanism and the fate of open strings.
\newblock {\em JHEP}, 01:019, 2001.

\bibitem{Kutasov:2000qp}
David Kutasov, Marcos Marino, and Gregory~W. Moore.
\newblock Some exact results on tachyon condensation in string field theory.
\newblock {\em JHEP}, 10:045, 2000.

\bibitem{Akhmedov:2001jq}
Emil~T. Akhmedov, Anton~A. Gerasimov, and Samson~L. Shatashvili.
\newblock On unification of rr couplings.
\newblock {\em JHEP}, 07:040, 2001.

\bibitem{Gerasimov:2001pg}
Anton~A. Gerasimov and Samson~L. Shatashvili.
\newblock On non-abelian structures in field theory of open strings.
\newblock {\em JHEP}, 06:066, 2001.

\bibitem{Kraus:2000nj}
Per Kraus and Finn Larsen.
\newblock Boundary string field theory of the dd-bar system.
\newblock {\em Phys. Rev.}, D63:106004, 2001.

\bibitem{Craps:2001jp}
Ben Craps, Per Kraus, and Finn Larsen.
\newblock Loop corrected tachyon condensation.
\newblock {\em JHEP}, 06:062, 2001.

\bibitem{Viswanathan:2001cs}
K.~S. Viswanathan and Y.~Yang.
\newblock Tachyon condensation and background independent superstring field
  theory.
\newblock {\em Phys. Rev.}, D64:106007, 2001.

\bibitem{Rashkov:2001pu}
R.~Rashkov, K.~S. Viswanathan, and Y.~Yang.
\newblock Background independent open string field theory with constant b field
  on the annulus.
\newblock 2001.

\bibitem{Alishahiha:2001tg}
Mohsen Alishahiha.
\newblock One-loop correction of the tachyon action in boundary superstring
  field theory.
\newblock {\em Phys. Lett.}, B510:285--294, 2001.

\bibitem{Andreev:2000yn}
Oleg Andreev.
\newblock Some computations of partition functions and tachyon potentials in
  background independent off-shell string theory.
\newblock {\em Nucl. Phys.}, B598:151--168, 2001.

\bibitem{Arutyunov:2001nz}
G.~Arutyunov, A.~Pankiewicz, and Jr. Stefanski, B.
\newblock Boundary superstring field theory annulus partition function in the
  presence of tachyons.
\newblock {\em JHEP}, 06:049, 2001.

\bibitem{deAlwis:2001hi}
S.~P. de~Alwis.
\newblock Boundary string field theory the boundary state formalism and d-brane
  tension.
\newblock {\em Phys. Lett.}, B505:215--221, 2001.

\bibitem{Bardakci:2001ck}
Korkut Bardakci and Anatoly Konechny.
\newblock Tachyon condensation in boundary string field theory at one loop.
\newblock 2001.

\bibitem{Lee:2001cs}
Taejin Lee.
\newblock Tachyon condensation and open string field theory.
\newblock {\em Phys. Lett.}, B520:385--390, 2001.

\bibitem{Lee:2001ey}
Taejin Lee.
\newblock Tachyon condensation, boundary state and noncommutative solitons.
\newblock {\em Phys. Rev.}, D64:106004, 2001.

\bibitem{Fujii:2001qp}
A.~Fujii and H.~Itoyama.
\newblock Some computation on g-function and disc partition function and
  boundary string field theory.
\newblock 2001.

\bibitem{DiVecchia:1999fx}
P.~Di~Vecchia and Antonella Liccardo.
\newblock D-branes in string theory. ii.
\newblock 1999.

\bibitem{Akhmedov:2001yh}
E.~T. Akhmedov, M.~Laidlaw, and G.~W. Semenoff.
\newblock On a modification of the boundary state formalism in off- shell
  string theory.
\newblock 2001.

\bibitem{Hsue:1970ra}
C.~S. Hsue, B.~Sakita, and M.~A. Virasoro.
\newblock Formulation of dual theory in terms of functional integrations.
\newblock {\em Phys. Rev.}, D2:2857--2868, 1970.

\bibitem{Fradkin:1985qd}
E.~S. Fradkin and A.~A. Tseytlin.
\newblock Nonlinear electrodynamics from quantized strings.
\newblock {\em Phys. Lett.}, B163:123, 1985.

\bibitem{Laidlaw:2000kb}
Mark Laidlaw.
\newblock Noncommutative geometry from string theory: Annulus corrections.
\newblock {\em JHEP}, 03:004, 2001.

\bibitem{Liu:1988nz}
Jun Liu and Joseph Polchinski.
\newblock Renormalization of the mobius volume.
\newblock {\em Phys. Lett.}, B203:39, 1988.

\bibitem{Polchinski:1998rq}
J.~Polchinski.
\newblock String theory. vol. 1: An introduction to the bosonic string.
\newblock Cambridge, UK: Univ. Pr. (1998) 402 p.

\end{thebibliography}

\end{document}